\documentstyle[prb,aps,epsf]{revtex}
\begin{document}

\twocolumn[\hsize\textwidth\columnwidth\hsize\csname@twocolumnfalse%
\endcsname

\title{Magnetic and quadrupolar order in a one-dimensional 
ferromagnet with cubic crystal-field anisotropy}

\author{M.\ Dudzi\'{n}ski$^{\dag}$, G.\ 
      F\'{a}th$^{\ddag}$\cite{byline1}, and J.\ Sznajd$^{\dag}$}
\address{\dag Institute for Low Temperature and Structure Research,
      Polish Academy of Sciences, 
      \protect\\ P.\ O.\ Box 1410, 50-950 Wroclaw 2, Poland\\
      \ddag Cavendish Laboratory, University of Cambridge, 
      Madingley Road, Cambridge CB3 0HE, England}

\date{\today}

\maketitle

\begin{abstract}
The zero temperature phase diagram of a one-dimensional $S=2$
Heisenberg ferromagnet with single-ion cubic anisotropy is studied
numerically using the density-matrix renormalization group method. 
Evidence is found that although the model does not involve 
quadrupolar 
couplings, there is a purely quadrupolar phase for large values of 
the anisotropy. The phase transition between the magnetic and
quadrupolar phases is continuous and it seems to be characterized 
by Ising critical exponents.
\end{abstract}


\vskip 0.3 truein
]


\section{Introduction}

It is well-known that crystal fields present in real materials
can influence their magnetic behavior considerably.\cite{CFbooks}
In this paper we study a ferromagnetic insulator,
with localized interacting magnetic moments, which are also 
subject to a crystal field of cubic symmetry. 
The finite-temperature critical 
phenomena in these systems have been investigated previously within
the Landau-Ginzburg-Wilson theory,\cite{AharonyKleinert} 
and the cubic universality class and its 
critical exponents have been analyzed. However, in such 
phenomenological
theories, the form of the order parameter is simply postulated,
using mainly symmetry arguments. To see whether such an
order really develops in the low-temperature regime one must start 
with the microscopic quantum Hamiltonian.

In terms of a quantum spin Hamiltonian, the magnetic moments 
are described by spin operators, and the crystal field takes the 
form of a single-site term. The lowest value of $S$
for which a cubic anisotropy term has any effect is $S=2$. 
Using this value our model Hamiltonian takes the form
\begin{equation}
  \label{model}
  H = - \sum_{\langle ij\rangle}  \bbox{S}_{i} \bbox{S}_{j} 
  - D \sum_{i} 
  \left[ (S^{x}_{i})^{4} + (S^{y}_{i})^{4} + 
  (S^{z}_{i})^{4} \right],
\end{equation}
with $S_{i}^{\alpha}$ ($\alpha=x,y,z$) denoting 
spin operators for $S=2$ at lattice site $i$.
The first term is the isotropic, 
ferromagnetic exchange interaction between nearest neighbors, 
and the second term represents the crystal field anisotropy;
for $S=2$ this is the most general single-site 
operator with cubic symmetry. The exchange term alone produces
ferromagnetic order in the low-temperature phase; the question
we want to address is how this order is affected by the cubic
crystal field.

The above Hamiltonian has been treated within the mean-field 
approximation (MFA) by many authors.\cite{SznajdValkova} It has
been found that the sign of the crystal-field constant~$D$ 
defines the easy axes of the magnetic ordering at low temperatures. 
For $D=0$ the system is isotropic and the spontaneous 
magnetization can lie along any direction; 
for $D < 0$ the cube diagonals $[111]$ are preferred;
for $0 < D < (4z)/3$, 
where $z$ is the number of nearest neighbors,
the directions parallel with the cube edges $[100]$ 
are chosen. Moreover, the MFA predicts that for 
$D > (4z)/3$ the system is disordered, 
irrespective of the temperature. 

Recently, two of us (M.D.\ and J.S.) have investigated the above 
model,\cite{OurRS}
applying the Raleyigh-Schr\"{o}dinger perturbation 
theory in the limit $D \rightarrow \infty$.
This calculation is valid for large $D$,
and the following picture emerges from it. In agreement
with the MFA, for large $D$, 
there is no magnetic order at any temperature,
i.e., the averages of the spin operators vanish:
\[
\langle S_{i}^{\alpha} \rangle = 0, 
\hspace*{1em} \text{for} \hspace*{1em}
\alpha=x,y,z.
\]
However, in contrast to the MFA prediction, 
the symmetry may still be spontaneously broken
by the appearance of a purely quadrupolar order, described by the 
averages
\[
\langle (S_{i}^{x})^2 \rangle \neq 
\langle (S_{i}^{y})^2 \rangle  =
\langle (S_{i}^{z})^2 \rangle.
\]
(One may interchange $x$, $y$, and $z$ in the above relation.)
This purely quadrupolar phase was predicted to 
exist in the model on a two-dimensional lattice at least at zero 
temperature. In three dimensions\cite{OurCB} 
and in higher dimensions
it should be present both in the ground state and at finite
(but low) temperatures. We suppose, that the quadrupolar phase also 
exists
in one dimension at zero temperature (note that the broken symmetry
is not continuous but discrete), although in this case the 
perturbation-theory arguments are weaker.

The result that there may appear a purely quadrupolar order in
the model in Eq.\ (\ref{model}), which contains only the bilinear,
``magnetic'' exchange and the single-site term, is rather
surprising. Usually, higher order exchanges are responsible for
quadrupolar ordering.\cite{SivardiereMorin} 
The appearance of quadrupolar order in 
the present case is entirely a quantum
effect; one would not obtain purely quadrupolar 
order by simply replacing the spin operators 
in Eq.\ (\ref{model}) with classical vectors,
and then seeking the configuration with minimum energy. 
The result is also of methodological interest. 
It is easy to understand that one cannot get purely 
quadrupolar order in this model by the MFA:
only magnetization can act as a mean field when the exchange 
term contains only first powers of the spin operators. 
Therefore, if the purely quadrupolar phase exists, 
we have the example that, in contrast with the general belief,
the MFA may give an answer which is qualitatively wrong
in high dimensions. 

Unfortunately, there is no fully reliable method which would enable 
us to investigate the model in two or three dimensions, for 
general $D$. Therefore, the aim of the present paper is to 
check and confirm the existence of the purely
quadrupolar phase numerically at least in the one-dimensional 
situation at zero
temperature. We use White's density-matrix renormalization group 
(DMRG) method,\cite{Whiteoriginal} which has a proven record of 
extreme efficiency in dealing with such problems. The method allows 
us to study the model for general $D \geq 0$, not only in the
limit of large $D$ as in Ref.\ \onlinecite{OurRS}, and we obtain 
the zero temperature phase diagram, together with some properties 
of 
the (quantum) phase transition separating the magnetic and 
quadrupolar 
phases. 

The paper is organized as follows: we begin with summarizing the 
information obtained from the MFA and the perturbation
theory in Sec.\ \ref{sec2}. Then we discuss how long-range order 
can be
detected using the DMRG in Sec.\ \ref{sec3}. Section \ref{sec4} is
devoted to the concrete numerical results, while Sec.\ V is a brief
discussion of our conclusions. We also attach an Appendix in which 
numerical predictions, obtained using identical techniques, are 
confronted with exact results in the test case of the 1D Ising 
model in a transverse field.

\section{Predictions from MFA \protect\\
and perturbation theory}
\label{sec2}

Let us first list the eigenvalues $e_{k}$ and eigenstates 
$| \psi_{k} \rangle$ of the single-site
term in the Hamiltonian in Eq.\ (\ref{model})
\begin{eqnarray}
e_{1}=-24 D   & \hspace*{2.5em} &  | \psi_{1} \rangle = 
 \frac{1}{\sqrt{2}} (| 2 \rangle + | -2 \rangle), \nonumber \\
e_{2}=-24 D   & \hspace*{2.5em} &  | \psi_{2} \rangle = 
 | 0 \rangle, \nonumber \\
e_{3}=-18 D   & \hspace*{2.5em} &  | \psi_{3} \rangle = 
 | 1 \rangle, \nonumber \\
e_{4}=-18 D   & \hspace*{2.5em} &  | \psi_{4} \rangle = 
 \frac{1}{\sqrt{2}} (| 2 \rangle - | -2 \rangle), \nonumber \\
e_{5}=-18 D   & \hspace*{2.5em} &  | \psi_{5} \rangle = 
 -| -1 \rangle,
\label{eigenstates}
\end{eqnarray}
where by $| 2 \rangle, | 1 \rangle, \ldots,  | -2 \rangle$ we 
denote
eigenstates of $S^{z}$. The five states are split into a doublet
and a triplet, and the doublet energy is lower for 
$D > 0$. The doublet is {\em nonmagnetic}, i.e.,
\begin{equation}
\label{nonmagneticproperty}
\langle \psi_{i} | S^{\alpha} | \psi_{j} \rangle = 0,
\end{equation}
for any $i,j=1,2$ and $\alpha=x,y,z$. 
Clearly, the crystal field with $D > 0$
opposes the magnetic ordering, since the nonmagnetic states
are preferred in the total wave function of the system.

For $D=0$ the model is isotropic and possesses the classical,
ferromagnetic ground state with saturated magnetization.
As mentioned in the Introduction, the MFA predicts 
\cite{SznajdValkova} that the magnetic phase exists for  
$0 < D < 8/3$ ($z=2$ for the chain), and that the directions of the 
cube edges $[100]$ are the easy axes in this phase. According 
to the MFA, the ground-state magnetization decreases with
$D$ and disappears continuously at $D=8/3$.  The MFA
result that the magnetic order vanishes at some finite $D$ is
a strong argument that this is indeed the case, since the MFA 
usually overestimates the tendency towards ordering. Nevertheless, 
the actual
value of $D$ at which the magnetic phase ends may be much
smaller than $8/3$ (for example, in the Ising chain in a
transverse field the MFA critical field is two times larger
than the exact critical field). Whether the actual phase
transition is continuous or discontinuous is also
a matter of question. We believe that the easy axes in the
magnetic phase are given correctly by the MFA. This approximation
neglects the quantum fluctuations in the ground state, but it
seems very unlikely that these could change the easy-axis 
directions,
especially in the vicinity of $D=0$, where the magnetization
is almost saturated and the quantum fluctuations are small.

Considering the large-$D$ limit now, we make use of the
perturbation theory developed in Ref.\ \onlinecite{OurRS}. Since
the two nonmagnetic states $| \psi_{1} \rangle$ and 
$| \psi_{2} \rangle$ dominate for large $D$, the model
reduces to an effective two-state Hamiltonian. Restricting our
attention to the one-dimensional
situation from now on, the effective
Hamiltonian $H_{\rm eff}$ can be written in terms of Pauli
matrices $\sigma_{i}^{\alpha}$ acting on the states
$| \psi_{1} \rangle$ and $| \psi_{2} \rangle$, and has the 
form\cite{OurRS}
%
%
\begin{eqnarray}
H_{\rm eff}   & =  
       & -\tilde{J_{1}} \sum_{i} 
          \left( \sigma_{i}^{x} \sigma_{i+1}^{x} + 
                 \sigma_{i}^{z} \sigma_{i+1}^{z} \right) 
            -\tilde{J_{2}} \sum_{i} 
           \left( \sigma_{i}^{y} \sigma_{i+1}^{y} \right) 
\nonumber\\
          && -\tilde{J_{3}} \sum_{i}
           \left( \sigma_{i}^{x} \sigma_{i+2}^{x} + 
                 \sigma_{i}^{z} \sigma_{i+2}^{z} \right) 
          -\tilde{J_{4}} \sum_{i}
          \left( 
            \sigma_{i}^{z} \sigma_{i+1}^{z} \sigma_{i+2}^{z} 
          \right. \nonumber\\
         && \left. \quad 
          - \sigma_{i}^{x} \sigma_{i+1}^{x} \sigma_{i+2}^{z} 
          - \sigma_{i}^{x} \sigma_{i+1}^{z} \sigma_{i+2}^{x} 
          - \sigma_{i}^{z} \sigma_{i+1}^{x} \sigma_{i+2}^{x} 
            \right)\nonumber\\
          &&+ {\cal O}(\frac{1}{D^4}),
\label{Heff}
\end{eqnarray}
\narrowtext
with
\begin{eqnarray*}
 \tilde{J_{1}} & = &  \frac{1}{2 D} -\frac{1}{12 D^{2}} 
                      +\frac{1}{72 D^{3}}
                      + {\cal O} \left( \frac{1}{D^{4}} \right),
                            \\
 \tilde{J_{2}} & = &  \frac{1}{8 D^{2}} 
                      +\frac{1}{32 D^{3}}
                      + {\cal O} \left( \frac{1}{D^{4}} \right), 
                      \\
 \tilde{J_{3}} & = &  \frac{1}{18 D^{3}}
                      + {\cal O} \left( \frac{1}{D^{4}} \right),
                           \\
 \tilde{J_{4}} & = &  \frac{1}{24 D^{3}}
                      + {\cal O} \left( \frac{1}{D^{4}} \right).   
\end{eqnarray*}
(The exchange coupling constant $J$ from Ref.\ \onlinecite{OurRS}
is set to $1$ in the present work.)
The quadrupole operators 
$[(S_{i}^{\alpha})^{2}-2]$ of the original model in Eq.\ 
(\ref{model}) 
turn out to be represented by linear combinations of Pauli matrices
as\cite{OurRS}
\begin{eqnarray}
[(S_{i}^{x})^{2}-2] \hspace*{0.8em}& \longmapsto &\hspace*{0.8em}
  \sqrt{3} \sigma_{i}^{x} - \sigma_{i}^{z}
   + {\cal O}(1/ D^2), \nonumber \\
\protect
[(S_{i}^{y})^{2}-2] \hspace*{0.8em}& \longmapsto &\hspace*{0.8em} 
  -\sqrt{3} \sigma_{i}^{x} - \sigma_{i}^{z}
   + {\cal O}(1/ D^2), \nonumber \\
\protect
[(S_{i}^{z})^{2}-2] \hspace*{0.8em}& \longmapsto &\hspace*{0.8em} 
  2 \sigma_{i}^{z}
   + {\cal O}(1/ D^2).
\label{Ssigma}
\end{eqnarray}
The analogous representations of $S_{i}^{\alpha}$ and 
$(S_{i}^{\alpha} S_{i}^{\beta} + S_{i}^{\beta} S_{i}^{\alpha})$,
where $ \alpha \neq \beta $, vanish up to ${\cal O}(1/D^{3})$.

In the lowest order in $1/ D$, only $\tilde{J_{1}}$
is present in the effective Hamiltonian, and $H_{\rm eff}$
describes the planar model with continuous SO(2) symmetry in the 
$XZ$ plane. While this model, conventionally written as the $XY$ 
model, is ordered at zero temperature in two dimensions
and above, in one dimension
it is known to be critical due to enhanced quantum 
fluctuations.\cite{XY,Bar-McC} The in-plane correlation function 
\[
\langle  \sigma_{i}^{x} \sigma_{i+l}^{x}  \rangle =
\langle  \sigma_{i}^{z} \sigma_{i+l}^{z}  \rangle
\]
decays asymptotically as a power law $1/\sqrt{l}$, where $l$ is 
the distance
between the two sites. Hence, on the basis of Eq.\ (\ref{Ssigma}),
we argue that for $D \rightarrow \infty$
a critical state, in which the quadrupolar correlation
function 
\[
\left\langle  \left[(S_{i}^{\alpha})^{2}-2\right]
\left[(S_{i+l}^{\alpha})^{2}-2\right]  \right\rangle, \qquad 
\alpha=x,y,z,
\]
decays asymptotically as $1/\sqrt{l}$, is approached. 

Taking into account the higher-order terms in $H_{\rm eff}$,
we obtain the planar, $XY$-type model with small perturbations.
The three-spin term with the coupling constant $\tilde{J_{4}}$
is crucial, because it reduces the symmetry of the effective 
Hamiltonian from the full rotational symmetry in the $XZ$ plane
to the symmetry of discrete rotations through angles
$\pm (2 \pi)/3$. In the one-dimensional case 
the $\tilde{J_{2}}$ and $\tilde{J_{3}}$ terms are
expected to be marginally irrelevant,\cite{Nomura} but in the lack 
of a
rigorous study of the off-diagonal three-point symmetry-breaking 
term with $\tilde{J_{4}}$,
we can only speculate (and then eventually check numerically) 
whether
the ${\cal O}(1/D^3)$ contributions are relevant or irrelevant. 
If they are relevant, they can stabilize the ``magnetic
order'' (i.e., the nonvanishing averages of $\sigma_{i}^{\alpha}$) 
in the effective model $H_{\rm eff}$, 
since the Mermin-Wagner and Coleman's theorems
do not hold when the symmetry is not continuous but discrete. 
If this is indeed the case, then Eq.\ (\ref{Ssigma}) implies 
that for large $D$ a quadrupolar
order, associated with the operators $[(S_{i}^{\alpha})^{2}-2]$,
exists in the original model of Eq.\ (\ref{model}) (see 
Ref.\ \onlinecite{OurRS} for details). Alternatively, if all the
perturbations turn out to be irrelevant, there should be a critical 
phase for large $D$. 

To summarize the above, we present the relevant (order) parameters 
to be analyzed as suggested by the MFA and the perturbation theory:
\begin{eqnarray}
m^{\alpha}  =  \langle S^{\alpha}_{i} \rangle, 
\label{defm}  \qquad
q^{\alpha}  =  \left\langle (S^{\alpha}_{i})^{2} - 2 \right\rangle.
\label{defq}
\end{eqnarray}
These ground-state averages are 
bulk values in the thermodynamic limit. Note, that in a disordered
state $m^{\alpha}=q^{\alpha}=0$ for $\alpha=x,y,z$.
Based on the above analysis we expect that for $0<D<D_m$ 
(where $D_m$ is probably smaller than the MFA value $8/3$) a 
magnetic
phase with
\begin{equation}
\label{magneticpar}
m^{x} \neq 0, \hspace*{1em} m^{y} = m^{z} = 0,
\hspace*{1.7em} q^{x} > q^{y} = q^{z},
\end{equation}
exists, while for $D>D_m$ a purely quadrupolar phase with
\begin{equation}
\label{quadrupolarpar}
m^{x} = m^{y} = m^{z} = 0, \hspace*{1.7em}
q^{x} > q^{y} = q^{z}
\end{equation}
emerges. (The indices $x,y,z$ in the above descriptions of the 
phases may be interchanged.)
In both phases the original cubic symmetry of the Hamiltonian is 
spontaneously broken. Note however that the subgroup remaining 
invariant is different in the two cases. As discussed above,
for large $D$ a critical (disordered) phase with 
$q^{\alpha}=0$ cannot be excluded either. In any case, the values
of $q^{\alpha}$ must go to zero for $D \rightarrow \infty$.

\section{The method}
\label{sec3}

The density-matrix renormalization group (DMRG) method is one of 
the
most reliable numerical techniques to study one-dimensional quantum
lattice problems. For a detailed description of the algorithm see 
Ref.\ \onlinecite{Whiteoriginal}. There are different 
implementations of the technique; nevertheless it seems that the
most accurate results can be
obtained by investigating finite systems with open boundary
conditions. In this setup the DMRG provides good approximations for 
the ground state and low-lying excited states of long, but finite 
quantum 
chains. This can then be supplemented by a detailed finite size
scaling analysis using appropriate scaling assumptions.

Usually for the case of spontaneous symmetry breaking the 
existence of long-range order is checked by studying 
appropriate two-point correlation functions $\langle A_n 
A_{n^\prime}\rangle$. The direct observation of one-point functions 
$\langle A_n \rangle$ is apparently hindered
by the fact that in finite systems, where
tunneling is always finite between different vacua, the ground 
state is
a symmetric combination of all possible ordering directions, and thus
naively $\langle A_n \rangle=0$. A possible way out, however, is
to break the symmetry artificially by a (small) auxiliary
field which then forces the system to make a choice between the 
{\it a priori}
undetermined directions, without noticably changing the value
of the order parameter. This method has considerable advantages in the
DMRG numerics where the measurement of two-point functions is rather
awkward due to the open boundary condition.

In our implementation the symmetry-breaking auxiliary 
fields are only applied to the first and last spins of the open 
chain. These two {\em boundary fields} are chosen to be identical, 
leaving the system symmetric for $i\to L-i$ reflections
($L$ is the chain length). Then the Hamiltonian which
is actually simulated in the numerical calculations is
\begin{equation}
   H \rightarrow H - {\bf h}\cdot \left( {\bf B}_1 +{\bf 
B}_L\right)
   \label{bf-setup}
\end{equation}
where ${\bf B}_i$ and ${\bf h}$ are vectors of appropriate 
single-site operators and boundary fields. Since the system is not
translation invariant, the one-point function $\langle A_n \rangle$
depends on the position $n$ near
the chain ends, but approaches its bulk value in the middle of the
system. If the system size is much larger than the correlation 
length
the one-point function rapidly saturates and forms a plateau.
Alternatively, if the chain length is too small comparing to the
correlation length (e.g., the system is critical or it is close to
criticality) no plateau appears to develop. 

There are two ways to estimate the order parameter and the 
associated correlation length from one-point functions. One can 
define the value $a_{L/2}\equiv \langle A_{L/2} \rangle$ 
measured in the middle of the 
chain and study its dependence on the total chain length $L$. 
It is expected that $a_{L/2}(L)\to a$ (with $a$ the bulk value 
of the order
parameter) exponentially fast as $L\to\infty$, whenever the system 
is away from criticality, while the convergence is only algebraic 
at 
criticality. A disadvantage of this method is that many independent 
DMRG runs must be done with different $L$ values, or one is forced 
to use the ``infinite lattice'' algorithm which is less precise. 

Alternatively, one can analyze the {\em profiles} 
$a_n\equiv \langle A_n \rangle$ as a function of
$n$ at a fixed length $L$, providing that $L$ exceeds the 
correlation
length $\xi$ reasonably (which can be checked {\it a posteriori}). 
In this latter case the profile is expected to have the following
asymptotic form far from the chain ends
\begin{equation}
   \label{fittingformula}
   a_n = a + c \frac{e^{-n / \xi}}{n^\chi} +
               c \frac{e^{-(L+1-n) / \xi}}{(L+1-n)^{\chi}} ,
\end{equation}
where $a$ is the bulk value of the order parameter, $\xi$ is the
associated correlation length, $\chi$ is a suitable exponent, and 
$c$ is a constant. In general one can assume that the 
order parameter and the correlation length measured this way 
are identical to their bulk values
derived from two-point correlation functions. However, the exponent 
of the polynomial prefactor $\chi$ is normally different from 
its bulk value and may depend on the boundary conditions chosen. 

In practice we analyze our numerical data by dividing the chain 
into smaller segments with a length
of the order of $\xi$, and perform local least-squares fits using 
the formula
in Eq.\ (\ref{fittingformula}). We have found it appropriate
to fix $\chi$ first and then look for its value which makes the 
other
fitting parameters the most stable in different segments as 
$n\to\infty$.
Our procedure of determining order parameters and correlation 
lengths was extensively tested on a simpler problem, the 1D Ising 
model in a transverse field (ITF), where exact answers to many 
questions of interest are available. We summarize our experience 
with this model in the Appendix.

In most cases the DMRG is very efficient to calculate spectral 
gaps. The method is less useful, however, if the ground state is 
(asymptotically) highly degenerate, since then several states 
have to be targeted together,
resulting in a rapid drop in accuracy.
In the cubic model under investigation we expect a 6-fold 
ground-state degeneracy in the magnetic phase, 
and a 3-fold degeneracy 
in the quadrupolar phase, in accordance with the number of 
easy-axis directions
for the two order parameters. The
6-fold degenerate case already exceeded our computational 
limitations, thus some of our conclusions with respect to 
that case are based on 
short chain exact diagonalization results only.

Errors in the quantitative predictions we report in the next 
sections stem from two sources, either from the limited accuracy 
of the numerical technique we
have been using, or from the approximate (asymptotic) nature of the 
formulas we applied for fitting and extrapolating the data. As for 
the DMRG errors, our results are somewhat more precise when the 
``finite 
lattice'' algorithm was used. In this case several 
iterations were done until convergence was reached. 
In principle, results obtained through the ``infinite
lattice'' method inevitably contain a systematic ``environment'' 
error, and thus should be treated with less confidence. However, 
in the present problem improvements due to the ``finite 
lattice'' iterations turned out to be rather small, 
providing enough ground to assume that even our ``infinite 
lattice'' 
results have sufficient accuracy.

In general we made various runs keeping different 
number of states $M$, and extrapolated our results for
$M\to\infty$, or at least checked that the results had been 
converged in
this parameter. Typically, we found good convergence in $M$
despite the fact that the maximum number of states kept did not 
exceed $M=85$. Note that the number of degrees of freedom
at a single site is five in the $S=2$ case, and the model in 
question has no continuous axial symmetry 
(total $S^z$ is not conserved)
which could have been utilized to facilitate the DMRG calculation 
in the usual way.

\section{Numerical results}
\label{sec4}

In order to obtain the zero temperature phase diagram and critical
properties of the model we carried out extensive numerical 
calculations
using exact diagonalization techniques on short chains and the DMRG
method on longer systems. In the case of open boundary conditions 
with
boundary fields the actual form of the Hamiltonian used in the
simulations was
\begin{eqnarray}
  \label{actmodel}
  H &=& - \sum_{i=1}^{L-1}  \bbox{S}_{i} \bbox{S}_{i+1} 
  - \sum_{i=1}^L D_i
  \left[ (S^{x}_{i})^{4} + (S^{y}_{i})^{4} + 
  (S^{z}_{i})^{4} \right] \nonumber\\ 
  &&-{\bf h}\cdot {\bf (B_1+B_L)}
\end{eqnarray}
with a reduced cubic crystal field at the chain ends (to 
counterbalance the missing bonds for the first and the last spins)
\begin{equation}
   D_i = \left\{ \begin{array}{ll}  D & {\rm if}\; i=2,\dots,L-1 \\
                            D/2 & {\rm if}\; i=1,L   \end{array} 
\right.
\end{equation}
and a general symmetry-breaking boundary field containing magnetic 
and quadrupolar terms
\begin{equation}
   {\bf B} = \left( \begin{array}{c}  
        S^x\\S^z\\(S^x)^2\\(S^y)^2\\(S^z)^2 \end{array} \right), 
\qquad
   {\bf h} = \left( \begin{array}{c}  
        h_x\\h_z\\h_{xx}\\h_{yy}\\h_{zz} \end{array} \right).
   \label{Bh}
\end{equation}
In order to avoid handling complex numbers, we did not include a 
boundary term proportional to $S^y$.

After the ground state had been found, we measured the  
order parameters. To facilitate further discussion we introduce the
notations
\begin{eqnarray}
m_{n}^{\alpha}  =  \langle S_{n}^{\alpha} \rangle, \qquad
q_{n}^{\alpha}  =  \langle (S_{n}^{\alpha})^{2}-2 \rangle
\label{defmq}
\end{eqnarray}
with $n=1,\dots,L$; $\alpha=x,y,z$.

\subsection{Ground state degeneracies and gaps}

\begin{figure}[hbt]
\epsfxsize=\columnwidth\epsfbox{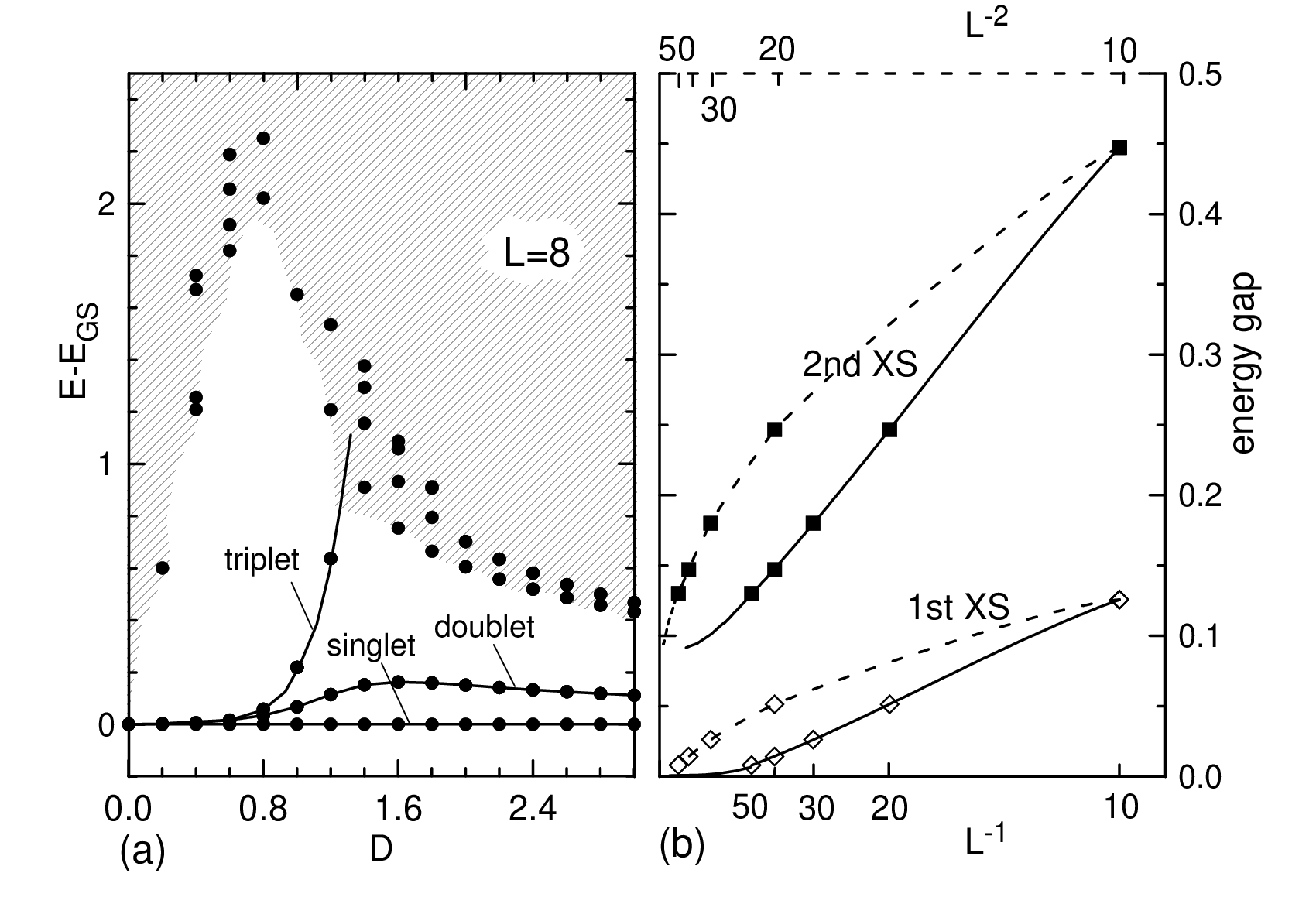} 
\caption{(a) Low-energy spectrum vs cubic anisotropy $D$ in a short 
chain with $L=8$ using periodic boundary condition. Shading 
schematically indicates higher-energy states. 
(b) Ground state--first excited state (singlet--doublet) gap (1st 
XS), 
and ground state--second excited state 
gap (2nd XS) as a function of $1/L$ (solid lines)
and $1/L^2$ (dashed lines) at $D=1.3$. 
The ticks in the horizontal axes correspond to $L=10,20,\ldots,50$.
\label{fig:degen} } 
\end{figure}

Figure \ref{fig:degen}(a) shows the lowest energy levels in a 
periodic chain with $L=8$ sites obtained using
exact diagonalization techniques. Although the chain is short, the 
6-fold degeneracy, consisted of a singlet, a doublet, and a triplet
is clearly discernible for small values of $D$. 
In the large $D$ (quadrupolar) region the degeneracy seems 
to be 3-fold, consisted of the singlet and the doublet only. In 
both regimes we observe signs of a gap above these states, although 
the number of data points ($L=4,6,8$) available for a given $D$ 
did not allow us to perform a detailed finite-size scaling study. 
Note that since the
broken symmetry is discrete we do not expect any massless Goldstone
modes in the ordered regimes. 
Excited states are rather massive domain walls, which 
separate domains with different ordering directions, and 
propagate in the system. 

As was discussed in the former section, we could only apply the 
DMRG for the 3-fold degenerate case. We calculated the low-lying 
four states in an open chain
at $D=1.3$, which is above the critical point $D_m$ (see later 
sections), by targeting four states without any boundary fields. 
The results, presented in Fig.\ \ref{fig:degen}(b), strongly 
support the 3-fold asymptotic degeneracy with
a finite gap $\Delta\approx 0.08\pm 0.02$ above it in 
the thermodynamic limit.
Unfortunately, this calculation was rather time consuming, 
which impeded us to repeat it for other $D$ values.

\subsection{Magnetic order}
\label{sec4B}

Let us now investigate the spontaneous magnetization and the
corresponding correlation length. In these calculations, purely
magnetic boundary fields were used, i.e.,  
$h_{xx}=h_{yy}=h_{zz}=0$ in Eq.\ (\ref{Bh}).

\begin{figure}[hbt]
\epsfxsize=\columnwidth\epsfbox{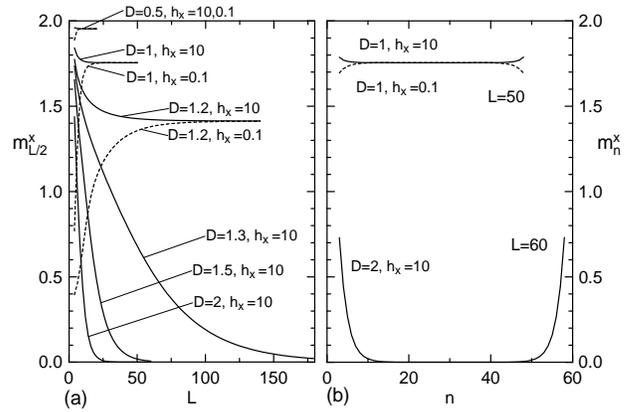} 
\caption{(a) The parameter $m^{x}_{L/2}$ vs the chain length $L$,
and (b) the profile $m^{x}_n$ vs the position $n$ at $L$
fixed, for some representative values of $D$. Purely
magnetic boundary fields applied along the $x$ axis as given by 
$h_{x}$.
\label{fig:magn} } 
\end{figure}

Expecting the magnetic easy axes along the cube edges we started 
with applying a boundary field along the $x$ axis setting
$h_{x}=10$ or $0.1$, $h_{z}=0$. Due to the symmetry, 
with such a boundary field one must obtain $m_{n}^{x} \neq 0$ 
and $m_{n}^{y}=m_{n}^{z}=0$. 
Figure \ref{fig:magn}(a) shows how the parameter $m_{L/2}^{x}$, 
measured in the center of the chain, changes with the total
chain length $L$ for representative values of the crystal-field 
strength $D$. One can see that for $D=0.5,1,1.2$ the parameter 
$m_{L/2}^{x}$ converges 
to finite values independent of $h_{x}$. This clearly indicates
a finite magnetic order for these $D$ values. On the other hand, 
for $D=1.3,1.5,2$ the parameter $m_{L/2}^{x}$ goes to zero as 
$L\to\infty$, meaning no magnetic order. 
As characteristic examples the magnetization profiles 
$m_{n}^{x}$ with $L$ fixed are shown in 
Fig.\ \ref{fig:magn}(b) for $D=1$ and $2$.
For $D=1$ a plateau corresponding to a finite spontaneous 
magnetization $m^{x}\approx 1.76$, while for $D=2$ a plateau 
with $m^{x}=0$ can be observed.

So far we have assumed that the cube edges give the easy axes in 
the magnetic phase. To verify this
assumption, we also applied boundary fields which do not
coincide with the expected easy directions. Choosing $h_{x}=0.04$ 
and $h_{z}=0.03$ we found that for $L\to\infty$ the parameter 
$m_{L/2}^{x}$ always tends to the same value 
as found above, while $m_{L/2}^{z}$ goes to zero, in agreement with
Eq.\ (\ref{magneticpar}).

\begin{figure}[hbt]
\epsfxsize=\columnwidth\epsfbox{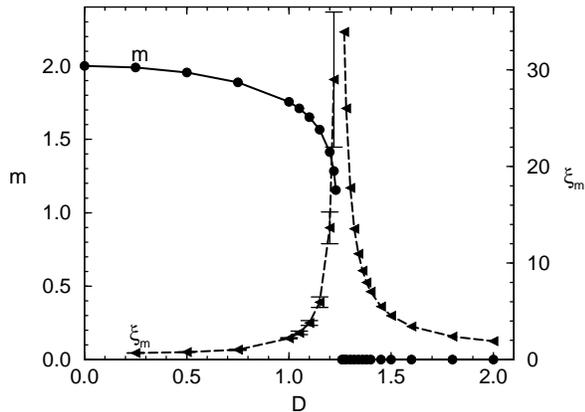} 
\caption{The spontaneous magnetization $m$ (left axis, dots) and 
the corresponding correlation length $\xi_{m}$ (right axis, 
triangles) as functions of $D$. 
Notice the large error bars for $\xi_{m}$ in the magnetic phase.
\label{fig:magnD} } 
\end{figure}

The above calculations allowed us to determine how the bulk
spontaneous magnetization $m$ changes with the crystal-field
strength $D$. We assume that monotonic dependencies of 
$m^{\alpha}_{L/2}$ on $L$, like those in Fig.\ \ref{fig:magn}(a),
provide correct bounds for $m$, as was found for the ITF chain 
in the Appendix. The magnetization $m$ is plotted in 
Fig.\ \ref{fig:magnD} as a function of 
$D$, together with the corresponding correlation length $\xi_m$,
discussed in details below. Since $\xi_m$ increases around the
phase transition, very long chains have to be studied in order
to find the large-$L$ limit of $m_{L/2}^{x}$ in this region. 
The largest $D$
for which we can see the magnetic order is $D=1.23$. For this
$D$, taking $L=400$, we obtain a finite $m$ with an uncertainty
of $0.003$ (for smaller $D$, shorter chains are sufficient to
calculate $m$ much more precisely). Due to the increasing 
correlation length, and thus computational limitations, 
we were not able to trace the dependence of
$m$ on $D$ further. Starting from $D=1.26$, the phase with no 
magnetic order is observed.

Although it cannot be shown directly that the spontaneous 
magnetization disappears continuously at the phase-transition
point, the sharp increase of the correlation length suggests
that the transition is continuous. We therefore attempt to fit
the dependence of $m$ on $D$, shown in Fig.\ \ref{fig:magnD}, 
assuming a power
law singularity and a linear regular term
\begin{equation}
\label{fitm}
m(D) = c_1 (D_m-D)^{\beta} + c_2 (D_m-D).
\end{equation}
The fit is very good and we obtain
\begin{eqnarray*}
D_m  =  1.2374 \pm 0.0004,\qquad \beta  =  0.127 \pm 0.004,   
\end{eqnarray*}
$c_1  =  2.15 \pm 0.05$, $c_2  =  -0.17  \pm 0.05$.
Interestingly, the exponent $\beta$ is very
close to that of the two-dimensional Ising model. The obtained
parameter $D_m$ is our best estimate of the phase-transition 
point.

To calculate the magnetic correlation length $\xi_{m}$, we analyze 
the profiles of $m_{n}^{x}$ at fixed $L$, analogous to those
presented in Fig.\ \ref{fig:magn}(b). As was described earlier, we
divide the chain into short sections and perform local fits
using the formula in  Eq.\ (\ref{fittingformula}). 

For each $D>D_m$, i.e., in the phase with no magnetic order,
the local values of the fitting parameters obtained with 
$\chi=0$ are very stable, and the convergence with increasing $n$
(towards the center of the chain) is convincing. The fitted order
parameter $m$ is always very close to zero. 
It seems that the decay of 
$m_{n}^{x}$ is asymptotically purely exponential, like the decay
of $m^{\rm ITF}_n$ in the disordered phase of the ITF chain, 
discussed in the Appendix. 
Due to the stability of local $\xi$'s, the correlation length 
$\xi_{m}$ in the phase with no magnetic order can be calculated
very precisely (at $D=1.27$ the error in $\xi_{m}$ is
around 0.1\% and it is smaller for larger $D$). 

In the magnetically ordered phase, for $D<D_m$, the local
fitting parameters obtained with $\chi=0$ are not stable. The 
local $\xi$'s increase towards the center of the chain, like in
the ordered phase of the ITF chain, suggesting the existence of
an algebraic prefactor. To calculate $\xi_m$, we vary $\chi$ and
look for its value where the local $\xi$'s are most stable. We 
do not find a universal prefactor, and the obtained $\chi$
depends on $D$ and $h_x$. However, the estimates of $\xi_m$ are 
independent of the boundary field, which justifies the procedure;
the uncertainties in $\xi_m$ are around $10$\%. (The same procedure
was tested for the ordered phase of the ITF chain and yielded
acceptable estimates of the correlation length --- see Appendix.)

The dependence of the correlation length $\xi_{m}$ on the 
crystal-field  strength $D$ is plotted in Fig.\ \ref{fig:magnD}. 
Due to the large error bars for $D<D_m$, the 
functional dependence can only be analyzed
for $D>D_m$. Here, assuming a second order transition again, we fit
the dependence of $\xi_{m}$ on $D$ with a power law
\begin{equation}
   \label{fitxim}
   \xi_{m} (D) = c_3 (D-D_m')^{-\nu}.
\end{equation}
The fit is again rather good, yielding the values
\begin{eqnarray*}
   D_m'  =  1.236 \pm 0.004,  \qquad   \nu  =  1.02 \pm  0.06.
\end{eqnarray*}
Thus, close to the phase transition both the spontaneous
magnetization and the corresponding correlation length 
seem to be well described by power laws. The two independent 
estimates of the phase-transition point, $D_m$ and $D_m'$ 
from Eqs.\ (\ref{fitm}) and (\ref{fitxim}), nicely coincide 
with each other. (In what follows, we will use $D_m$, which 
is more precise.) Moreover,
the exponents $\beta$ and $\nu$ are close to those of the
two-dimensional Ising model. Altogether, we have found strong
indications that, as far as the magnetic quantities $m$ and 
$\xi_{m}$
are concerned, the phase transition is continuous and 
is described by the critical exponents of the two-dimensional
Ising model.

\subsection{Quadrupolar order}
\label{sec4C}

We now investigate the
quadrupolar order, associated with the parameter $q^{\alpha}$
defined in Eq.\ (\ref{defq}). While the quadrupolar order must be
present in the magnetic phase ($D<D_m$), the question whether
it exists in the phase with no magnetic order ($D>D_m$) is
of central interest. 

Expecting the quadrupolar easy axes along the cube edges,
as given in Eqs.\ (\ref{magneticpar}) and (\ref{quadrupolarpar}),
we began with applying boundary fields along the $x$ axis. These
were purely quadrupolar: 
$h_{\alpha}=0$; $h_{xx} \neq 0$, $h_{yy}=h_{zz}=0$ in Eq.\ 
(\ref{Bh}); or purely magnetic:
$h_{x} \neq 0$, $h_{z}=0$; $h_{\alpha \alpha}=0$. With such 
boundary fields, due to the symmetry, one must obtain
$q_n^x \neq q_n^y=q_n^z$.

\begin{figure}[hbt]
\epsfxsize=\columnwidth\epsfbox{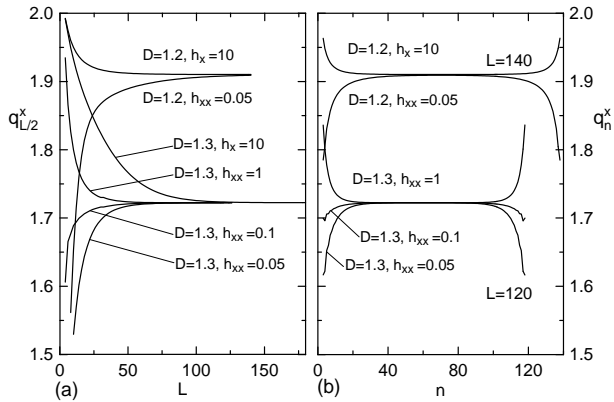} 
\caption{(a) The parameter $q^{x}_{L/2}$ vs the chain length $L$,
and (b) the profile $q^{x}_n$ vs the position $n$ at
$L$ fixed, for two values of $D$.
Purely magnetic ($h_{x}$) and purely quadrupolar ($h_{xx}$)
boundary fields applied along the $x$ axis.
\label{fig:quad} } 
\end{figure}

We found that
in both phases the order parameter $q_{L/2}^{x}$ measured in
the chain center converges with increasing $L$ to finite
values which do not depend on the actual boundary fields. 
Figure \ref{fig:quad}(a) shows 
these dependencies for $D=1.2<D_m$ and $D=1.3>D_m$.
The appropriate profiles of $q^{x}_{n}$ with 
$L$ fixed are plotted in Fig.\ \ref{fig:quad}(b). 
In both phases the picture is
typical for an ordered state: $q^{x}_{n}$  has plateaus
in the middle of the chain, which correspond to finite bulk
$q^{\alpha}$. Thus, we find quadrupolar order on both sides of 
$D_m$,
confirming the existence of a purely quadrupolar phase for $D>D_m$.

To verify that the cube edges give the easy axes of the quadrupolar
ordering, we applied a
quadrupolar boundary field having all three components 
different from zero: $h_{\alpha}=0$, $h_{xx}=0.03$, $h_{yy}=0.02$,
and $h_{zz}=0.01$. Then, in both phases, it was observed that
$q^{x}_{L/2}>q^{y}_{L/2}=q^{z}_{L/2}$ in the large-$L$ limit,
and the limiting values of $q^{\alpha}_{L/2}$ were the same as
found above, in accordance with Eqs. (\ref{magneticpar}) 
and (\ref{quadrupolarpar}). 

\begin{figure}[hbt]
\epsfxsize=\columnwidth\epsfbox{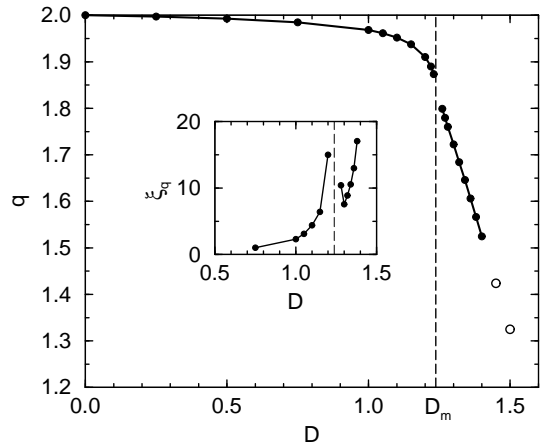} 
\caption{The quadrupolar order parameter $q$ as a function of $D$. 
For $D=1.45$ and $1.5$ upper bounds on $q$ are depicted as the 
open circles. The inset shows very rough estimates of the
corresponding correlation length $\xi_{q}$.
\label{fig:quadD} } 
\end{figure}

Assuming that monotonic dependencies of $q^{\alpha}_{L/2}$ on $L$ 
provide correct bounds for the bulk value $q$, we determined  
$q$ for various values of $D$. 
($q$ is defined as the largest of $q^{\alpha}$ in a 
broken-symmetry state.) The results are 
depicted as bold points in Fig.\ \ref{fig:quadD}; the precision
in $q$ is $10^{-3}$ or better. We observed that,
starting from $D \approx 1.3$, the associated correlation length
$\xi_q$ increases rapidly with $D$ 
(see the inset in Fig.\ \ref{fig:quadD}). 
At the same time, 
the DMRG precision for $q^{\alpha}_{n}$ fell dramatically. 
For these reasons, already at $D=1.45$ (and for larger $D$ too) 
we were unable to obtain precise results
for a chain long enough to show us the large-$L$ limit of 
$q^{\alpha}_{L/2}$ (for $D=1.45$ and $1.5$ we obtained upper bounds 
on $q$, which are shown as open circles in Fig.\ \ref{fig:quadD}).
We had similar problems close to the transition point $D_m$.

The effective Hamiltonian $H_{\rm eff}$ in Eq.\ (\ref{Heff})
may help to understand the difficulties for $D\agt 1.4$. Notice, 
that the symmetry-breaking term $\tilde{J_4}$ in $H_{\rm eff}$ is
very small, $\tilde{J_4}/\tilde{J_1}={\cal O}(1/D^2)$. 
Thus even if $\tilde{J_4}$ is relevant, the correlation 
length is expected to diverge rather rapidly as $D\to\infty$. 
On the other hand, the observed abrupt increase in $\xi_{q}$
and the emerging
numerical difficulties beyond $D\sim 1.4$ may indicate an 
additional phase transition in which the quadrupolar long-range 
order vanishes, and gives rise to an extended critical phase
with algebraically decaying quadrupolar correlations 
beyond a critical
value $D_q$. A naive linear fit to the data points for $q$ in Fig.\ 
\ref{fig:quadD} would yield a value $D_q\approx 2.1$, but we are 
not in a position to give any firm claims in this question. 
Note however that if $D_q=\infty$, 
i.e., if the quadrupolar phase extends 
to $D\to\infty$, the curve $q(D)$ should finally bend upwards, and
in the data points in Fig.\ \ref{fig:quadD},
together with the upper bounds at $D=1.45$ and $1.5$,
the (small) curvature is consistently downwards.

Now we analyze the dependence of $q$ on $D$ in Fig.\ 
\ref{fig:quadD}
close to the phase-transition point $D_m$.
For $D<D_m$ we assume a power-law singularity and a linear
regular term
\begin{equation}
\label{fitqleft}
q(D) = q_{\rm left} + c_4 (D_m-D)^{\theta} + c_5 (D_m-D).
\end{equation}
Fitting yields
\begin{eqnarray*}
q_{\rm left}  =  1.838 \pm 0.01,  \qquad \theta =  0.5 \pm 0.1,    
\end{eqnarray*}
$c_4   =  0.44 \pm 0.1$, $c_5   =  -0.36 \pm 0.1$.
Thus, on this side of the phase-transition point, the quadrupolar
order parameter appears to have singular behavior, but the exponent
$\theta$ differs from $\beta$ describing the spontaneous
magnetization.

For $D>D_m$, we suppose the behavior is regular, which is
suggested by Fig.\ \ref{fig:quadD}. Our fitting formula is now
\begin{equation}
\label{fitqright}
q(D) = q_{\rm right} + c_6 (D-D_m) + c_7 (D-D_m)^{2},
\end{equation}
giving
\begin{eqnarray*}
q_{\rm right}   & = & 1.840 \pm 0.002,  
\end{eqnarray*}
$c_6     =  -1.83 \pm 0.03$, $c_7     =  -0.6  \pm 0.1$.
The left and right estimates of the order parameter 
$q$ at $D_m$ are in agreement, which supports our scaling 
assumptions.
However, we cannot exclude the possibility that a singular 
derivative 
in $q$ exists also for $D>D_m$, so close to $D_m$ that we
do not observe it in our numerics.

To calculate the correlation length $\xi_{q}$, we analyze
the profiles of $q^{\alpha}_{n}$, like those in Fig.\ 
\ref{fig:quad}(b). In our opinion,
only a purely quadrupolar boundary field is suitable for this
purpose. With a magnetic boundary field, the decay of 
$q_{n}^{\alpha}$ could be driven by the decay of $m_{n}^{\alpha}$,
and so governed by $\xi_{m}$ and not $\xi_{q}$. 

Using the formula in Eq.\ (\ref{fittingformula}) again, we find 
that 
both for $D<D_m$ and $D>D_m$ the prefactor exponent $\chi$ is
nonzero and it depends on $D$ and the boundary field. The fitting
yields rough estimates to $\xi_q$ (Fig.\ \ref{fig:quadD} inset),
the errors are believed to be around $20$\%.
The results suggest that $\xi_{m} = \xi_{q}$ for 
$D<D_m$, which is the expected answer, since the magnetic and
quadrupolar order is intimately connected in this phase.
For $D>D_m$, the correlation length $\xi_q$ increases with $D$
starting from $D\approx 1.3$, as mentioned above, but, due to the 
poor precision, we cannot resolve whether $\xi_{q}$ diverges
(numerically it increases sharply) for $D \rightarrow D_m$
or it remains finite.

\section{Conclusions}

In summary, we have studied the ground-state properties of a
one-dimensional $S=2$ ferromagnetic spin chain with single-site 
cubic crystal-field anisotropy $D$. 
We argued that in contrast with the mean-field
prediction, perturbation theory suggests 
the possibility that a purely quadrupolar phase exists for
large values of $D$. This quadrupolar phase was expected to be 
separated
from the magnetic phase by a continuous quantum phase transition at 
a critical point $D_m$.
We verified this conjecture by investigating the model
numerically using the density-matrix renormalization group method 
on open chains with special, symmetry-breaking boundary conditions.

In most cases, the method provided precise estimates of the
magnetic and quadrupolar order parameters. Very close to the 
phase-transition point and in the large-$D$ limit, however, 
our results were 
less accurate due to the rapidly increasing correlation lengths.
For the correlation lengths themselves we could only obtain rather 
rough estimates. 

Evidence has been obtained that, in qualitative agreement with the
mean-field prediction, the spontaneous magnetization diminishes
continuously at the phase-transition point. Regarding the magnetic
properties, the transition seems to be characterized by the 
critical exponents of the two-dimensional Ising model.
This could be plausible, since the extra 
broken symmetry of the magnetic ground state with respect 
to that of the quadrupolar ground state is just a $Z_2$ 
subgroup --- the same group as the one breaking down spontaneously 
in the case of the Ising model.

Both in the magnetic and nonmagnetic phases
we demonstrated the presence of a quadrupolar order.
In the former case the quadrupolar order is just a 
``secondary'' effect,
inevitably present in any $S\ge 1$ model with magnetic order.
In the latter case, however, it is the ``primary'' order,
constituting a qualitatively different broken-symmetry phase.
At the phase-transition point $D_m$, the quadrupolar order 
parameter is continuous. For $D\alt D_m$, we could clearly 
discern a singularity in the derivative of the order parameter, 
and found that the quadrupolar correlation length diverges together 
with the magnetic correlation length. 
For $D\agt D_m$, our observations are much less concrete: the 
order parameter can be fitted reasonably well by a low-order 
(regular) polynomial without any singular terms, and,
although the quadrupolar correlation length
increases as $D_m$ is approached from above, our results
do not suggest unambiguously a divergence on 
this side. It is not clear for us whether the phase transition 
involving the magnetization should have any precursor in the 
quadrupolar fluctuations above $D_m$.

Mainly due to computational limitations we were unable to resolve 
convincingly the question whether the quadrupolar phase extends 
to $D\to \infty$ or there is a finite value $D_q$ where 
the quadrupolar
long-range order disappears. This question is unique to the
one-dimensional situation --- in two dimensions 
and higher we expect a finite 
quadrupolar order at zero temperature for any $D>D_m$.

Finally, the scenario for the magnetic-to-quadrupolar phase 
transition 
was supported by the analysis of the degeneracy of the ground state 
as a function of $D$. We observed a 6-fold and a 3-fold asymptotic 
degeneracy below and above $D_m$ respectively, 
in accordance with the
the number of possible ordering directions in the two phases.

The confirmation of the purely quadrupolar phase in the 
one-dimensional 
case gives rise to the belief that such a phase
should also exist in higher dimensions at sufficiently low 
temperatures
when the cubic crystal field is strong. Increasing temperature
necessarily destroys the quadrupolar order and leads to a 
finite-temperature phase transition between the quadrupolar 
and a completely disordered phase. Note that the 
correct critical theory of this 
transition cannot be obtained by simply substituting spins with
classical vectors in our model, 
since this approach is unable to account for the 
purely quadrupolar order. 
The usual practice of neglecting quantum fluctuations by treating 
spins classically around a finite-temperature phase transition
would confront fundamental difficulties in this case.

\acknowledgements
The authors thank Steven White for the suggestion of applying
symmetry-breaking boundary conditions. Valuable discussions with
Enrico Carlon, Andrzej Drzewi\'{n}ski, and Uli Schollw\"{o}ck
are also acknowledged. This research was supported by 
the KBN (Poland) grant no.\ 2P03B07214 
and the EPSRC (UK) grant no.\ GR/L55346.

\begin{appendix}
\section*{A test calculation}

Our numerical procedure and extrapolation methods were tested on 
the zero-temperature 1D Ising model in a transverse field (ITF) 
where an exact solution for the bulk quantities
is available.\cite{ITF,Bar-McC} Our experience with this model
provided useful guidelines to refine our
procedure in the study of the cubic model of Eq.\ (\ref{model}).

The Hamiltonian of the ITF model on an open chain with length $L$ 
is defined as
\begin{eqnarray}
   \label{HITF}
   H^{\rm ITF} &=& - \sum_{i=1}^{L-1} \sigma_{i}^{z} 
\sigma_{i+1}^{z}
                 - \sum_{i=1}^L \Gamma_i
                    \sigma_{i}^{x}\\
   \Gamma_i &=& \left\{ \begin{array}{ll} \Gamma & {\rm if}\; 
i=2,\dots,L-1 \\
                            \Gamma/2 & {\rm if}\; i=1,L   
\end{array}
   \right. \nonumber
\end{eqnarray}
where $\sigma_{i}^{\alpha}$ denotes Pauli matrices at site $i$, 
and $\Gamma$ is the transverse magnetic field. Note that in the 
above
definition the spins at the chain ends, being coupled to one 
neighbor only,
are subject to the reduced field $\Gamma/2$.
We concentrate on the spontaneous magnetization
$ m^{\rm ITF} = \langle \sigma_{i}^{z} \rangle $
and the corresponding correlation length $\xi^{\rm ITF}$.
For $\Gamma=0$ the Hamiltonian $H^{\rm ITF}$ has the classical,
ferromagnetic ground state with $m^{\rm ITF} = \pm 1$, and
the single-site term in $H^{\rm ITF}$ tends
to destroy the spontaneous magnetization; in this respect the
ITF model is similar to the cubic model in Eq.\ (\ref{model}). 
In the thermodynamic limit there are two phases in the ITF chain,
distinguished by the order parameter $m^{\rm ITF}$. For 
$0 < \Gamma < 1$ the ground state is ordered due to spontaneous 
symmetry
breaking 
\begin{equation}
   m^{\rm ITF} = \pm \left(1-\Gamma^{2}\right)^{1/8}, 
\hspace*{2em}  
\xi^{\rm ITF} = - \frac{1}{2 \log \Gamma},
\label{orderedITF}                       
\end{equation}
while for $\Gamma>1$ there is a disordered phase with
\begin{equation}
   m^{\rm ITF} = 0, 
\hspace*{2em}  
\xi^{\rm ITF} = \frac{1}{\log \Gamma}.
\label{disorderedITF}                       
\end{equation}
At $\Gamma=1$ the ground state is critical and $\xi^{\rm ITF} = 
\infty$.

In our study the following values of the transverse field were 
considered: 
$\Gamma=0.9$ (ordered phase) where $\xi^{\rm ITF} \approx 4.75$; 
$\Gamma=1.1$ (disordered phase) where $\xi^{\rm ITF} \approx 
10.49$; 
and the critical point $\Gamma=1$ with $\xi^{\rm ITF}=\infty$.
We computed the position dependent order parameter 
$m^{\rm ITF}_{n} = \langle \sigma_{n}^{z} \rangle$ 
using the DMRG method with boundary fields up to $L=200$. 
As a symmetry-breaking boundary field we considered
\begin{equation}
   B = \sigma^z
\end{equation}
in Eq.\ (\ref{bf-setup}), and we used two different values of $h$: 
a strong $h=10$ and a weak $h=0.1$ boundary field.
In the case of the ITF chain
the DMRG is extremely precise even close to 
the critical point.\cite{Gaborandcousin} Keeping a relatively small 
number of states $M=64$ the
truncation errors are already negligible and the fitting procedure 
can be tested on numerically exact data. We emphasize that although 
rigorous results are available for quantities in 
the thermodynamic limit, the
behavior of relatively short chains with complicated boundary 
conditions is not feasible for a study with purely 
analytical tools even for the ITF model.

\begin{figure}[hbt]
\epsfxsize=\columnwidth\epsfbox{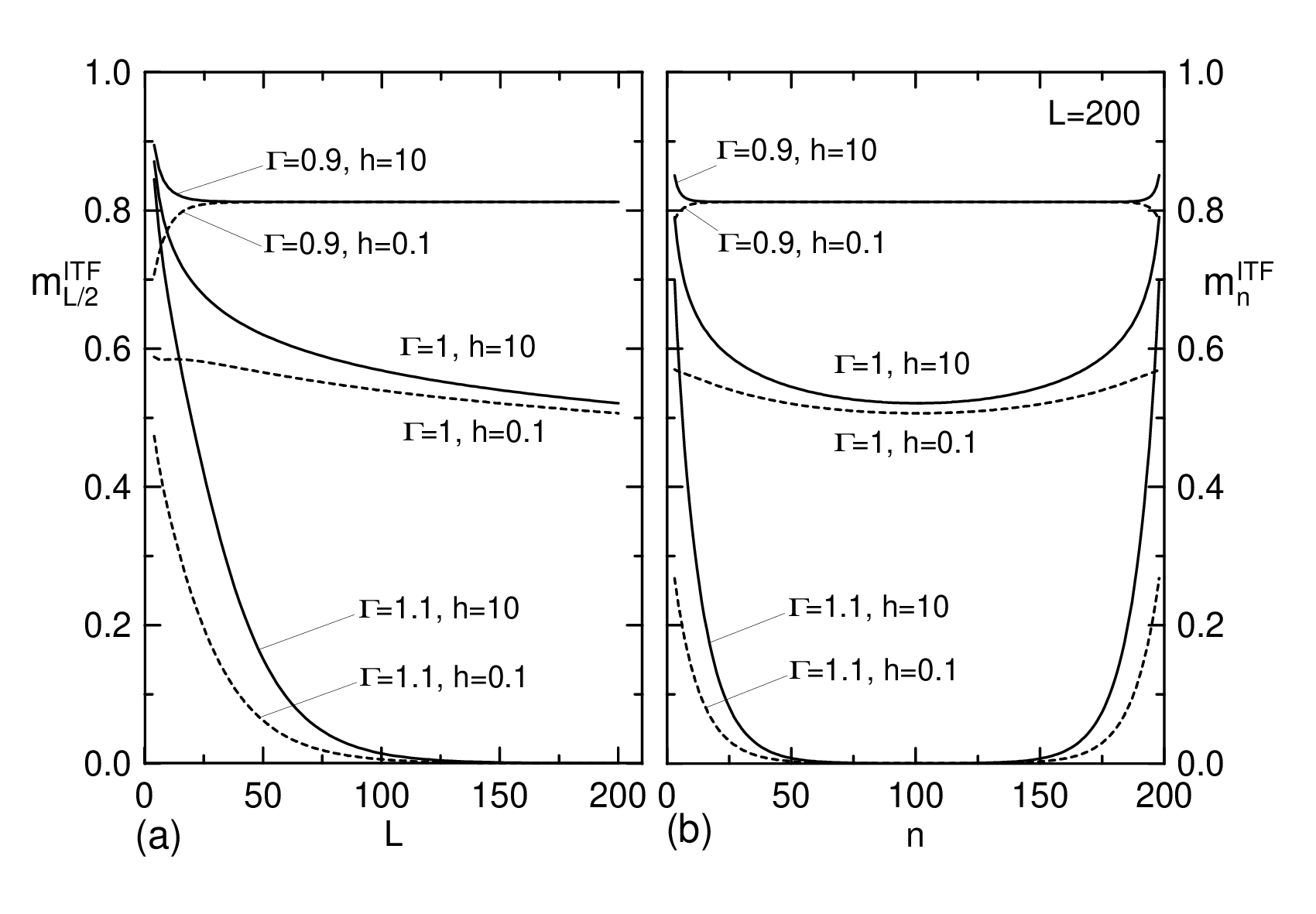} 
\caption{ITF chain results. (a) Order parameter $m^{\rm 
ITF}_{L/2}$ measured in the center of the chain vs the chain length 
$L$. (b) Order parameter profiles $m^{\rm ITF}_{n}$ as a function 
of position in the chain $n$ at fixed $L=200$. 
Curves are labeled by the values of the
transverse field $\Gamma$ and the boundary-field strength $h$.
\label{fig:ITF} } 
\end{figure}

The value of the order parameter at one of the central spins
$m^{\rm ITF}_{L/2}$ as a function of the chain length $L$ is
shown in Fig.\ \ref{fig:ITF}(a). For these data the 
``infinite-lattice''
DMRG was used from $L=4$ to $L=200$, thus a systematic 
``environment error'' cannot be excluded.\cite{Gaborandcousin}
In the ordered phase, $m^{\rm ITF}_{L/2}$ converges to the 
same finite value for both values of $h$, and the two dependencies 
are monotonic. At each $L$, correct bounds for the bulk 
$m^{\rm ITF}$ in Eq.\ (\ref{orderedITF}) are obtained,
and at $L=200$, ten digits of $m^{\rm ITF}$ are recovered.
In the disordered phase, $m^{\rm ITF}_{L/2}$ converges
to zero for both boundary fields $h$. At the critical point, 
the limit of $m^{\rm ITF}_{L/2}$ does not
show up in Fig.\ \ref{fig:ITF}(a), we can only say that the 
correlation length must be of the order of $L=200$ or larger.

In Fig.\ \ref{fig:ITF}(b) we show the order parameter profiles 
$m^{\rm ITF}_{n}$ for a chain length fixed at $L=200$. This length
exceeds the exact correlation lengths $\xi^{\rm ITF}$ for 
$\Gamma=0.9$ 
and $\Gamma=1.1$ many times. 
There are plateaus for $\Gamma=0.9$ and $\Gamma=1.1$,
which correspond to the bulk values of the order parameter.
There is no plateau at the critical point where the correlation 
length
is infinite and the dependence of the profile is algebraic.

In order to calculate the correlation length $\xi^{\rm ITF}$,
we analyze the profiles from Fig.\ \ref{fig:ITF}(b) using the 
ansatz in 
Eq.\ (\ref{fittingformula}). As is described in Sec.\ III we first 
fix a value of the exponent $\chi$ and then fit for the
other parameters. Repeating the fits in different sections of the 
chain we seek the value of $\chi$ which makes the other 
parameters the most stable as a function of $n$.

For $\Gamma=1.1$ the value of the exponent which yields the most 
stable fitting parameters is $\chi \approx 0$. 
From positions $n=50$ to $n=100$, and for both $h$, the fitted
correlation length $\xi$ agrees with the exact $\xi^{\rm ITF}$ with 
a precision of 7 digits! It seems that, asymptotically,
the decay of $m^{\rm ITF}_{n}$ in the disordered phase is 
purely exponential. This should be compared with the exponent
$\chi_{\rm bulk}=1/2$ characterizing the decay of the bulk 
two-point correlation function in this phase.\cite{Bar-McC} 
(For a similar difference in algebraic prefactors of one 
and two-point correlation functions in 
an $S=1$ chain see Ref.\ \onlinecite{Sor-Aff}.)

\begin{figure}[hbt]
\epsfxsize=\columnwidth\epsfbox{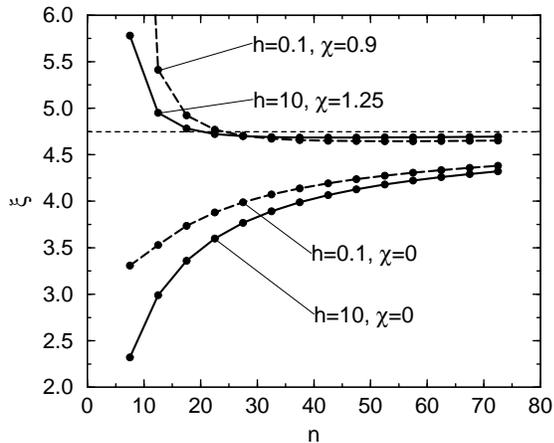} 
\caption{ITF chain results. Ordered phase $\Gamma=0.9$. The local 
values of the correlation length $\xi$ from Eq.\ 
(\protect\ref{fittingformula}), obtained
with different boundary fields $h$ and 
exponents $\chi$, as a function of the position $n$ at $L=200$.
The exact correlation length $\xi^{ITF}$ is shown as a 
horizontal line.
\label{fig:ITF2} } 
\end{figure}

In the ordered phase for $\Gamma=0.9$ we can also start with
$\chi = 0$. In this case, however, the fitting parameters are not 
stable, 
and $\xi$ increases with $n$ for both values of $h$, as is 
illustrated by Fig.\ \ref{fig:ITF2}.
This suggests that a nonzero prefactor is present. The value of 
$\chi$ which makes the correlation length $\xi$ the most stable 
seems to depend on the boundary field applied. We obtain
$\chi \approx 1.25 \pm 0.05$ for $h=10$ and $\chi \approx 0.9 \pm 
0.05$ for $h=0.1$, we cannot see a universal prefactor. (Note 
that for the bulk two-point functions the exponent of the 
prefactor is $\chi_{\rm bulk}=2$ in this phase.) The values
of $\xi$ calculated with the above prefactors are shown
in Fig.\ \ref{fig:ITF2}. It is seen that using these prefactors we 
find better estimates for $\xi^{\rm ITF}$ than with $\chi=0$.

Anticipating computational limitations in the investigation
of the $S=2$ cubic model in Eq.\ (\ref{model}),
we also analized shorter chains, such that 
$L \approx 10 \xi^{\rm ITF}$ during the test calculations. 
For $\Gamma=1.1$ and $L=100$
four digits of the exact $\xi^{\rm ITF}$ are recovered;
for $\Gamma=0.9$ and $L=50$ only rough estimates for 
$\xi^{\rm ITF}$ are obtained, the error is around 10 \%.

Our observations can be summarized as follows:
\begin{itemize}
\item
The order parameter $m^{\rm ITF}$ can be calculated accurately
in both phases;
it can be decided whether the system is ordered or not.
\item
In the ordered phase, upper and lower bounds for 
$m^{\rm ITF}$ are obtained by considering strong 
and weak boundary fields, respectively.
\item
The correlation length $\xi^{\rm ITF}$ 
can be calculated very precisely in the disordered phase.
\item
In the ordered phase, 
rather rough estimates of $\xi^{\rm ITF}$ can be found.
\end{itemize}

\end{appendix}


\end{document}